%% file: main.tex
\documentclass[12pt,a4paper]{elsarticle}
\usepackage{ulem}
\usepackage{lineno,hyperref}
\modulolinenumbers[5]

\journal{XXX}

\usepackage{amssymb}
\usepackage{amsmath}
\usepackage{amsfonts}
\usepackage{amsthm}
\usepackage{wasysym}
\usepackage{bbold}
\usepackage{bigints}

\usepackage[ansinew]{inputenc}
\usepackage[T1]{fontenc}
\usepackage{lmodern}
\usepackage{textcomp}
\usepackage{graphicx}

\usepackage[]{color}

\usepackage{fancyhdr}
\usepackage{algorithm}

\usepackage{algpseudocode}

\usepackage{tcolorbox}
\tcbuselibrary{breakable}
\usepackage{caption}
\usepackage{subcaption}

\newcommand{\conditional}[2]{\left\vert\vphantom{\frac{1}{1}}\right.}

\theoremstyle{definition}
\newtheorem{definition}{Definition}

\newtcolorbox{algbox}[3][]{
  width=\textwidth,colframe=black,colback=white,
  sharp corners, breakable,
  before={\captionof{algorithm}{#2}\label{#3}},
  #1
}

\makeatletter
\renewcommand{\ALG@name}{Procedure}
\makeatother

\begin{document}

\begin{frontmatter}

\title{On the enumeration of leaf-labelled increasing trees with arbitrary node-degree}

\author{Johannes Wirtz\corref{mycorrespondingauthor}}
\cortext[mycorrespondingauthor]{Corresponding author}
\ead{jwirtz@lirmm.fr}

\begin{abstract}
We consider the counting problem of the number of \textit{leaf-labeled increasing trees}, where internal nodes may have an arbitrary number of descendants. The set of all such trees is a discrete representation of the genealogies obtained under certain population-genetical models such as multiple-merger coalescents. While the combinatorics of the binary trees among those are well understood, for the number of all trees only an approximate asymptotic formula is known. In this work, we validate this formula up to constant terms and compare the asymptotic behaviour of the number of all leaf-labelled increasing trees to that of binary, ternary and quaternary trees.
\end{abstract}

\end{frontmatter}

\section{Introduction}
\input{intro}
\section{Preliminaries}
\input{preliminaries}
\section{Enumeration of labelled increasing trees}
\input{result-ode}
\section{Labelled increasing trees of restricted node-degree}
\input{restricted-degree-ode}

\section{Discussion}
\input{discussion}

\section*{Acknowledgements}
I thank Filippo Disanto for helpful comments concerning the methodology of analytic combinatorics. I also thank Vaclav Kotesovec for stimulating discussions on the counting sequence \texttt{A256006} and its asymptotic formula.\\
This work was financially supported by the \href{https://dfg.de}{DFG} through their Walter-Benjamin programme (WI 5589/1-1).

\bibliography{references}
\end{document}

%% file: intro.tex
Trees are among the most important data structures in mathematical biology and used to model and interpret various evolutionary processes \cite{semple:phylo,gillespie:popgen}. In this work we want to study a counting problem concerning a type of tree encountered in population-genetical models of the Cannings type \cite{cannings:model} and their infinite-population limits (e.g. the $\Lambda$- and other multiple-merger coalescents \cite{moehle:coalescent,sagitov:coalescent,pitman:coalescent,berestycki:beta}) or spatial models \cite{barton:slfv}. In population genetics, the interpretation  of a tree is that of a genealogy \cite{wakeley:coaltheory}; the leaves are regarded as members of some population that has evolved over time, with the branches of the tree pointing backward in time, and internal nodes denoting common ancestors of the members; hence, the internal nodes determine how closely related individuals are in relation to each other. Branch lengths represent the time that one needs to go back into the past until a common ancestor is found.\\
Of course, trees can also be interpreted like that with a phylogenetical application in mind (see e.g. \cite{steel:shape}). In contrast to population genetics, the assumption that the leaves are "alive" at the same point in time is usually not of such big importance when studying relationships in between species; still, the notion of most recent common ancestors and the interpretation of branch lengths as evolutionary times certainly can be carried over to this field.\\
Taking the population size $2N$ to infinity in the Wright-Fisher model \cite{wright:evolution}, we recover Kingman's coalescent \cite{kingman:coalescent}. Under this process, the genealogy $\gamma$ of a sample of some finite size $n\in\mathbb{N}_{\geq 2}$ has an elegant stochastical representation: It is always structured as a binary tree, and the times at which internal nodes occur in the past (i.e., times at which two lineages find a common ancestor) are determined by independent exponentially distributed random variables. Hence, they occur almost surely at distinct points in time, and one can discretize $\gamma$ by labeling the internal nodes by integers from $1$ to $n-1$ according to their order of appearance. That way, exact branch lengths are discarded, but otherwise the tree $\tau$ that is obtained from $\gamma$ contains the same genealogical information. However, $\tau$ is a purely combinatorial object and there are only finitely many possibilities for such a tree on $n$ leaves, so one might be interested in the number of distinct trees or in one of their statistics \cite{aldous:balance,fischer:balance}.\\
When addressing  questions like these, it is necessary to specify when one tree $\tau_1$ "equals" another tree $\tau_2$. For instance, one could restrict oneself to considering only planar equivalence classes, but one could also regard an "entangled" version of some tree as a new object. Similarly, one could consider the leaves as indistinguishable from each other, or as pairwise different, which can, for example, be realized by labeling them. The first is in accordance with the way individuals are viewed under the Wright-Fisher model (and Kingman's coalescent), because individuals produce exact copies of themselves which cannot be told apart from the original. However, Kingman's coalescent can be quite easily equipped with a mutation process that can indeed cause all individuals in a sample to be of their own distinct "genotype", which is reflected more accurately in the second scenario. An overview of the most frequently used definitions can be found in \cite{murthag:dendrograms}\\
In the context of Kingman's coalescent there are two concepts that have proven convenient. The first one is to employ a variant of the yule process \cite{yule:process}, where $\tau$ is built up by successive splits, with the child nodes appended below the parent node and one child being the "right" and the other the "left" child (see e.g. \cite{wirtz:emg}). The tree $\tau$ is an "increasing binary tree" as defined in \cite{flajolet:combinatorics} (\textit{pp}.144); often, these trees are just called "coalescent trees" \cite{aldous:probability,disanto:yulenodeimbalance,disanto:external}. Let the set of all such trees with $n$ leaves be denoted by $\mathcal{T}^{(2)}_n$. Importantly, two trees $\tau,\tau'\in\mathcal{T}^{(2)}_n$ are not necessarily equal if they are isomorphic as graphs (for instance, exchanging the subtrees of $\tau$ below the root in general results in a different tree). The cardinality $T^{(2)}_n=|\mathcal{T}^{(2)}_n|$ is $(n-1)!$.\\
The second concept is to consider the leaves as carrying distinct labels $\alpha_1$-$\alpha_n$, and, for each $i=n-1,\dots,1$ generate an internal node with label $i$ that becomes the father node of two currently "free" nodes. Here, we distinguish trees by their isomorphy classes (although a "label-preserving" isomorhpy; see Definition~\ref{def:isom}). On the other hand, exchanging the labels $\alpha_j,\alpha_k$ of two nodes in general creates a different tree. We call these trees \textit{leaf-labeled increasing trees} (to be in conformity with both  \cite{murthag:dendrograms,bondini:trees}) and denote the set of all such trees on $n$ leaves by $\mathcal{L}^{(2)}_n$, noting that several other names have been used throughout the literature for this combinatorial class \cite{murthag:dendrograms,aldous:cladograms,wirtz:tld}. The number of different trees of size $n$ is given by $L^{(2)}_n=|\mathcal{L}^{(2)}_n|=n!(n-1)!2^{-n+1}$ \cite{frank:dendrograms}.\\
$\mathcal{T}^{(2)}_n$ and $\mathcal{L}^{(2)}_n$ are the "natural" discretizations of Kingman's coalescent, because it induces the uniform distribution on those trees \cite{aldous:probability,wirtz:emg}. In models which may produce non-binary genealogies (focusing somewhat on the $\Lambda$-coalescents, of which Kingman's coalescent is a special case), when asking which tree class is suitable to describe those genealogies, it is therefore logical to consider variants of $\mathcal{T}^{(2)}_n$ and $\mathcal{L}^{(2)}_n$ where the mechanisms are modified accordingly. We will for now let the resulting classes be denoted by $\mathcal{T}_n$ and $\mathcal{L}_n$, postponing formal definitions. We are particularly interested in their cardinalities.\\
\begin{figure}
\includegraphics[scale=.75]{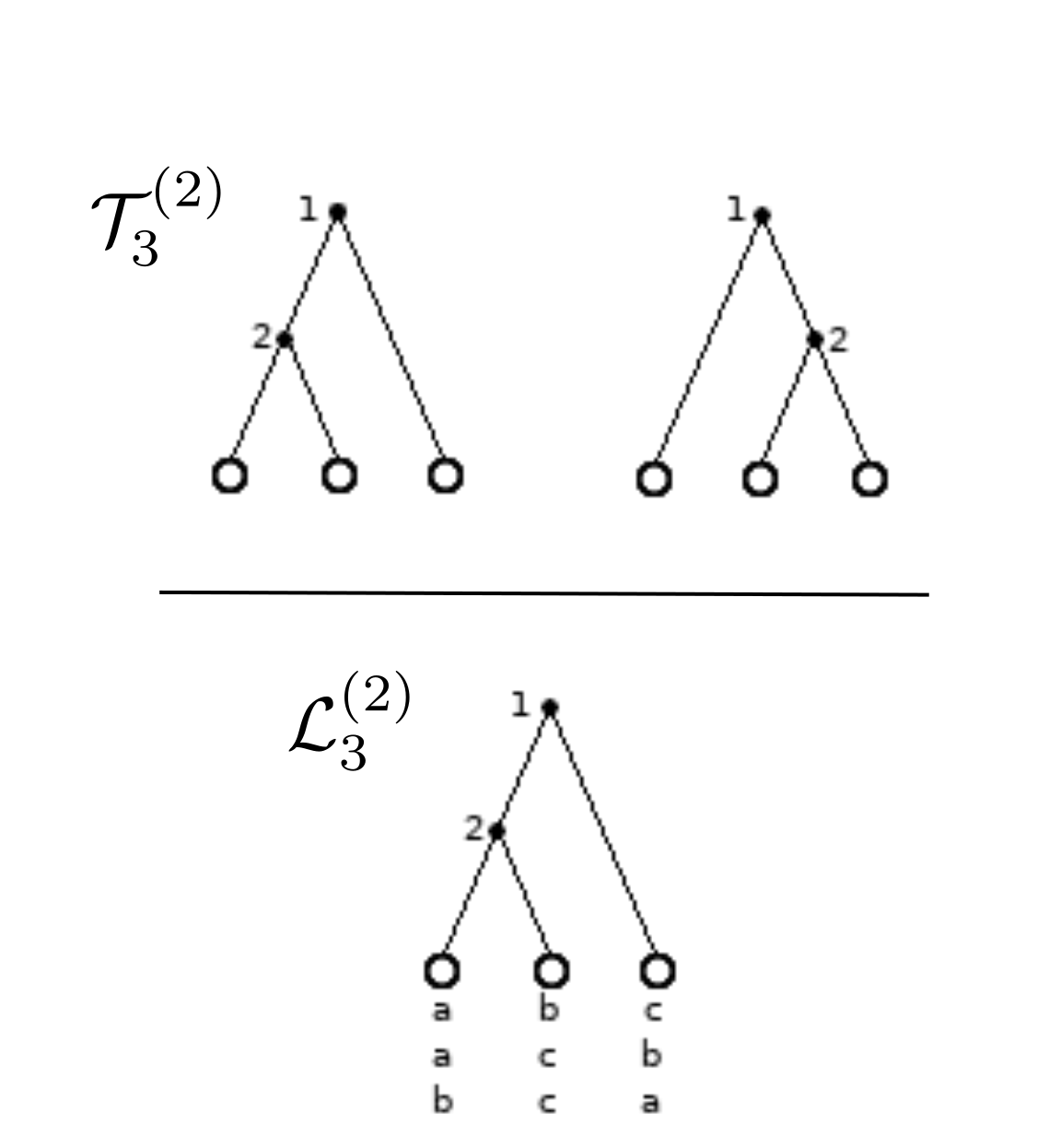}
\includegraphics[scale=.75]{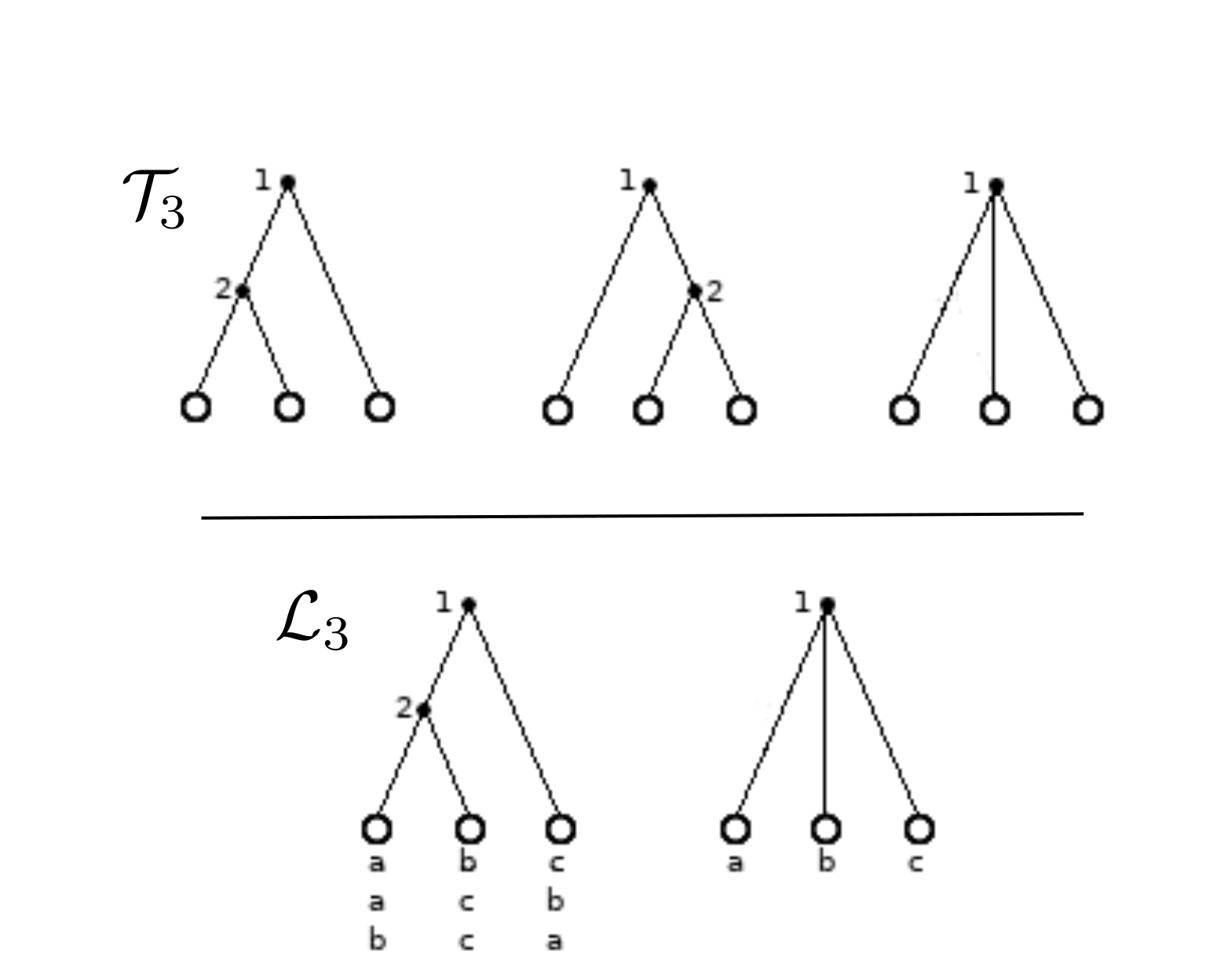}
\caption{\textit{Left:} All binary trees on $3$ leaves under the respective definitions of $\mathcal{T}^{(2)}_3$ and $\mathcal{L}^{(2)}_3$; \textit{right}: Allowing higher outdegree of nodes adds new topologies to $\mathcal{T}_3$ and $\mathcal{L}_3$ compared to their binary versions.}
\end{figure}
The set $\mathcal{T}_n$ that represents the "arbitrary-outdegree" extension of $\mathcal{T}^{(2)}_n$ is the set of \textit{increasing Schroeder trees} \cite{schroeder:trees}, and the number of different trees on $n$ leaves is given by $|\mathcal{T}_n|=n!/2$. It is rather surprising that this correspondence has apparently gone unnoticed until recently \cite{bondini:trees}. In the latter work, a number of properties of $\mathcal{T}_n$ regarding height and balance are derived as well.\\
Regarding $\mathcal{L}_n$, which extends binary leaf-labeled trees to general leaf-labeled trees, no closed formula for the cardinality is known. Their counting sequence $1,1,4,29,336,5687, 132294$ has identifier \texttt{A256006} in the Online Encyclopedia of Integer Sequences (\href{https://oeis.org/}{\texttt{OEIS}},\cite{oeis}), and the asymptotic formula
\begin{equation}\label{eq:kotesovec}
\mathcal{L}_n \sim c_{\infty} n^{2n+8/3} 2^{-n} \exp(-2n)
\end{equation}
has been proposed and experimentally validated \cite{kotesovec:litrees}, with $c_{\infty}\approx 4.001655$. The main purpose of this work is to give a mathematical proof of the correctness of the exponential and polynomial factors in this formula. This will be achieved by applying techniques from analytic combinatorics. Afterwards, we will compare the asyptotic behaviours of the numbers of trees that exist when outdegree restrictions are not completely dropped, but the admissible "arity" of internal nodes is gradually increased from $2$ to $\infty$.

%% file: preliminaries.tex
A tree is a connected and acyclic graph $\tau=(V,E)$ on a set of vertices $V$ (\textit{nodes}) and $E\subseteq V\times V$;  edges are directed. Additionally, we assume that there is no node in $\tau$ with indegree of more than $1$, so that every node has at most one unique direct predecessor. For $|V|\geq2$, since $\tau$ is connected, this means that there are nodes of outdegree $0$ which are called terminal nodes or \textit{leaves}; other nodes are called internal nodes. We also assume that $\tau$ is \textit{rooted} (there is a unique \textit{root} node $r\in V$ of indegree $0$ "preceding" all other nodes). We demand also that all internal nodes have outdegree $> 1$, so there are no "redundant" nodes. Finally, we assume a labelling as described in the introduction (compare also \cite{murthag:dendrograms,bondini:trees}).
\begin{definition}
A \textit{leaf-labeled increasing tree} $\tau$ of size $n$, $n\in \mathbb{N}_{\geq 1}$ (with \textit{leaf-labels} $\{\alpha_1,\dots,\alpha_n\}$) is a rooted directed tree with the following properties:
\begin{enumerate}
\item There are $n$ leaves, each carrying one of the distinct labels $\alpha_1,\dots,\alpha_n$.
\item If $n\geq 2$, the $m$ internal nodes ($m\leq n-1$) are labeled by $1$-$m$, where the root $r$ carries the label $1$, the sequence of labels on any path from $r$ to a leaf in $\tau$ is monotonously increasing, and every number is used precisely once.
\end{enumerate}
For short, we will refer to such trees simply as \textit{labeled increasing trees}.
\end{definition}
The labeled increasing tree of size $1$ is a single leaf labeled $\alpha_1$.\\
Any labeled increasing tree $\tau$ can be generated by executing the following procedure: Start with $n$ nodes labeled $\alpha_1,\dots,\alpha_n$. As long as the graph is not connected, create a node with label $i$, where $i$ is the smallest natural number $\geq 1$ that has not yet been used, and make this node the predecessor of an arbitrary number $k\geq 2$ of existing nodes. Finally, when after adding $1\leq m\leq n-1$ nodes the graph is connected, relabel the internal nodes by exchanging the label $m$ with $1$, $m-1$ with $2$ and so on. If the above procedure takes exactly $n-1$ iterations, all internal nodes in $\tau$ have outdegree $2$, and we call $\tau$ \textit{binary}. In the same way, a labeled increasing tree whose internal nodes have outdegree up to $3$ ($4$,$K\geq 4$) is called \textit{ternary} (\textit{quaternary}, $K$\textit{-ary}).\\
The interpretation of this in the context of the Wright-Fisher model is as described in the introduction. The leaves are a collection ("sample") of individuals alive in the present, and, going back in time, we encounter coalescent events (represented by internal nodes) at which individuals or their predecessors find a common ancestor. Their order of appearance in time is determined by their labels. Other possible interpretations of a labeled increasing tree $\tau$ include that of a "merger history", where the leaves represent firms or enterprises, and the internal nodes determine the order in which they merge into one \cite{orrick:mergers}.\\
\begin{definition}\label{def:isom}
 Given two labeled trees $\tau_1=(V_1,E_1), \tau_2=(V_2,E_2)$, a \textit{label-preserving isomorphism} is a bijective function $\phi:V_{\tau_1}\rightarrow V_{\tau_2}$ such that
\begin{itemize}
\item $\forall v_1,v_2\in V_1 $: $v_1$ is a predecessor of $v_2$ $\Leftrightarrow$ $\phi(v_1)$ is a predecessor of $\phi(v_2)$
\item For all $v_1\in V_1$, it holds that the label of $v_1$ is the same as that of $\phi(v_1)$
\end{itemize}
\end{definition}
We say that two labeled increasing trees $\tau_1,\tau_2$ are \textit{equal} if we can construct a label-preserving isomorphism between them, and \textit{distinct} if such an isomorphism does not exist. 
\begin{figure}
\includegraphics[scale=1.8]{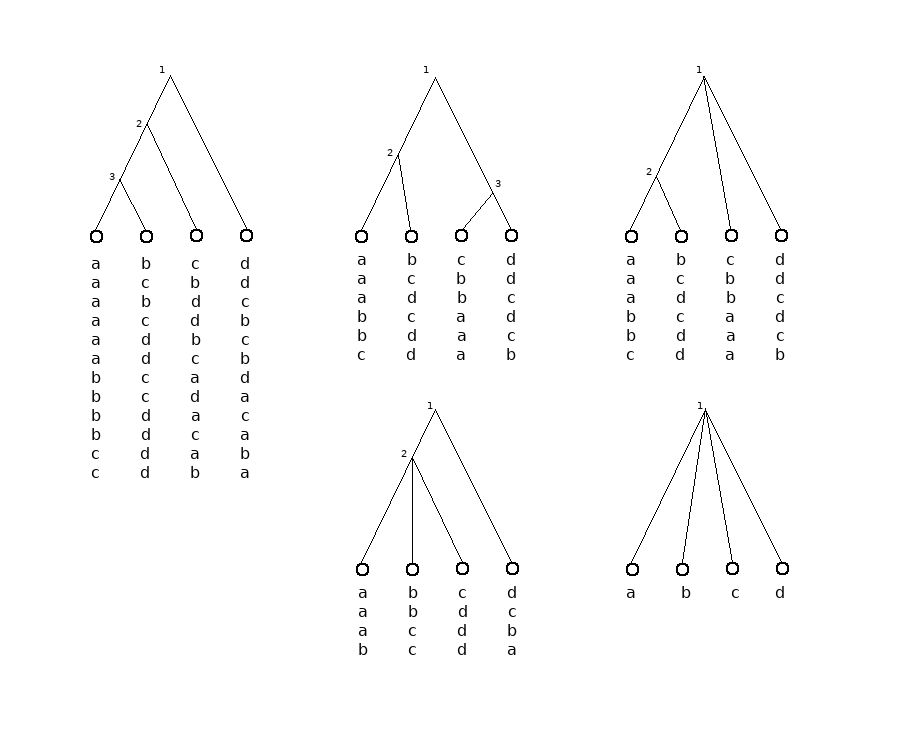}
\caption{All $29$ trees of $\mathcal{L}_4$, leaf labels represented by $\{a,b,c,d\}$}
\end{figure}
Finally, we introduce some notation:
\begin{definition}
\hfill\\
\begin{itemize}
\item For $n\in\mathbb{N}{\geq 1}$, let $\mathcal{L}_n$ denote the set of all distinct labeled increasing trees $\tau$ of size $n$ with leaf-labels $\{\alpha_1,\dots,\alpha_n\}$.
\item For $n\in\mathbb{N}{\geq 1}$, let $L_n$ denote the number of distinct labeled increasing trees $\tau$ of size $n$, i.e. $L_n=|\mathcal{L}_n|$
\end{itemize}
Concerning outdegree restrictions, we may write $\mathcal{L}^{(2)}_n$ (resp. $\mathcal{L}^{(K)}_n$) to denote the set of binary ($K$-ary) labeled increasing trees, and likewise, $L^{(2)}_n$ ($L^{(K	)}_n$) to denote their numbers.
\end{definition}

%% file: result-ode.tex
\subsection{The counting sequence and its generating function}
For any $n\in\mathbb{N}_{\geq 2}$ the following holds for $L_n$:

\begin{equation}L_n=\binom{n}{2}L_{n-1}+\binom{n}{3}L_{n-2}+\dots+\binom{n}{n-2}L_3+\binom{n}{n-1}L_2+\binom{n}{n}L_1\end{equation}

because we can generate any tree on $n$ leaves by choosing $k$ leaves to join below a new node labeled by $1$, then generating a tree from $\mathcal{L}_{n-k+1}$ out of the remaining free nodes. Let $\tilde{l}_n:=L_n/n!$. Then for $n\geq 2$,\\

\begin{equation}\tilde{l}_n=\frac{n-1}{2!}\tilde{l}_{n-1}+\frac{n-2}{3!}\tilde{l}_{n-2}+\dots+\frac{3}{(n-2)!}\tilde{l}_{3}+\frac{2}{(n-1)!}\tilde{l}_{2}+\frac{1}{n!}\tilde{l}_{1}\end{equation}

The exponential generating function $\tilde{L}(z):=\sum_{n\geq 1}\tilde{l}_n z^n$ of the $L_n$ can thus be expressed by

\begin{align}
\sum_{n\geq 1}\tilde{l}_n z^n &=\sum_{n\geq 2}\left(\sum_{k=2}^n\frac{n-k+1}{k!}\tilde{l}_{n-k+1}\right) z^n + z\\
&=\sum_{n\geq 2}\left(\sum_{k=2}^n\frac{n-k+1}{k!}\tilde{l}_{n-k+1}z^{n-k}z^k\right) + z\\
&=\sum_{n\geq 1}n\tilde{l}_{n}z^{n-1}\left(\sum_{k=2}^\infty\frac{1}{k!}z^k\right) + z
\end{align}

From this, it follows that the generating function must solve the differential equation
\begin{equation}
\tilde{L}(z)=\tilde{L}'(z)\left(\exp(z)-1-z\right)+z
\end{equation}
However, the solution to this equation is singular at $0$ and therefore not amenable to the usual methods of analytic combinatorics. Taking things one step further and dividing by a double factorial leads to a situation which can be understood better. Define $l_n:=L_n/(n!(n-1)!)$. Then for $n\geq 2$,

\begin{eqnarray}l_n=\frac{1}{2!}l_{n-1}+\frac{1}{3!(n-1)}l_{n-2}+\frac{1}{4!(n-1)(n-2)}l_{n-3}\\
+\dots+\frac{3!}{(n-2)!(n-1)!}L_{3}+\frac{2!}{(n-1)!(n-1)!}l_{2}+\frac{1}{n!(n-1)!}l_{1}\end{eqnarray}

The \textit{doubly-exponential} generating function $L(z):=\sum_{n\geq 1}l_n z^n$ takes the form

\begin{equation}
\sum_{n\geq 1}l_n z^n =\sum_{n\geq 1}\left(\sum_{k=2}^n\frac{(n-k+1)!}{k!(n-1)!}l_{n-k+1}\right) z^n + z
\end{equation}

It can be reformulated as an "infinite" integral equation, namely

\begin{equation}\label{eq:degf}
L(z)=z+\frac{z}{2}L(z)+\sum_{k=3}^{\infty}\frac{1}{k!}\underbrace{\int\dots\int}_{k-2} L(z)\underbrace{\mathrm{d}z\dots\mathrm{d}z}_{k-2}\cdot z
\end{equation}
where we let $\underbrace{\int\dots\int}_{j} L(z)\underbrace{\mathrm{d}z\dots\mathrm{d}z}_{j}$ denote the $j$th antiderivative of $L(z)$ whose first $j+1$ Taylor coefficients are all zero.
\subsection{Proof of the main result}
We will consider truncated versions of this equation (say, with $k\leq K$). We can transform the truncated integral equation into a differential equation by performing multiple derivations, and then use the methodology discussed in Ch.VII of \cite{flajolet:combinatorics} to determine coefficient asymptotics.\\
First, note that $\underbrace{\int\dots\int}_{j} U(z)\underbrace{\mathrm{d}z\dots\mathrm{d}z}_{j}\cdot z\leq \int U(z)\mathrm{d}z\cdot z^{j}$ for $j\in\mathbb{N}_{\geq 1}$ and for a function $U(z)$ analytic on a neighborhood of $0$ with positive Taylor coefficients. Because of that, we know that a solution to
\begin{align}\label{eq:upper}
\notag u_K(z)=z&+\frac{z}{2}u_K(z)+\sum_{k=3}^{K}\frac{1}{k!}\underbrace{\int\dots\int}_{k-2} u_K(z)\underbrace{\mathrm{d}z\dots\mathrm{d}z}_{k-2}\cdot z\\
&+\frac{\exp(z)-\sum_{j\leq K}z^j/j!}{z^K}\underbrace{\int\dots\int}_{K-1} u_K(z)\underbrace{\mathrm{d}z\dots\mathrm{d}z}_{K-1}
\end{align}
with $u_K(0)=0$ is an upper bound to $L(z)$. Importantly, $\frac{\exp(z)-\sum_{j\leq K}z^j/j!}{z^K}$ can be represented by the analytic function $=\sum_{j>K}z^{j-K}/j!$. Likewise, a function that solves
\begin{equation}\label{eq:lower}
s_K(z)=z+\frac{z}{2}s_K(z)+\sum_{k=3}^{K}\frac{1}{k!}\underbrace{\int\dots\int}_{k-2} s_K(z)\underbrace{\mathrm{d}z\dots\mathrm{d}z}_{k-2}\cdot z
\end{equation}
is a lower bound.\\
For example, ignoring all $k\geq 4$ in Equation~\eqref{eq:degf} leads to
\begin{equation}\label{eq:2nddeg}
s_3(z)=z+\frac{z}{2}s_3(z)+\frac{z}{6}\int s_3(z) \mathrm{d}z
\end{equation}
corresponding to the differential equation
\begin{equation}\label{eq:d2nddeg}
s_3''(z)-s_3'(z)\cdot\frac{1+1/6z}{1-z/2}-\frac{1}{3(1-z/2)}s_3(z)=0
\end{equation}
We are going to explicitly derive the coefficient asymptotics using the aforementioned method for this differential equation. Equation \eqref{eq:d2nddeg} has a singular point at $2$. The quantity $\delta_1$ refers to the \textit{residue} $\lim_{z\rightarrow 2}(z-2)\cdot\frac{1+1/6z}{1-z/2}=\frac{8}{3}$ of the coefficient of the first-order term in that equation, and likewise $\delta_2=\lim_{z\rightarrow 2}(z-2)^2\cdot\frac{1}{3}=0$ denotes the square residue of the term of order $0$. The \textit{indicial equation} is 
\begin{equation}
\theta(\theta-1)+\frac{8}{3}\theta=0
\end{equation}
with roots $0$ and $-\frac{5}{3}$. Applying Theorem VII.10 from \cite{flajolet:combinatorics}, the coefficient asymptotics of $L(z)$ are determined, up to a constant, by the formula $n^{2/3}2^{-n}$.\\
One can see quite easily by induction that truncating at an arbitrary $K>3$ always leads to the same result in that the asymptotic term of highest order for the coefficients of $s_K$ is always one of $\mathcal{O}n^{2/3}2^{-n}$. In fact, the smallest root of the indicial equation, i.e. the one determining the polynomial term in the asymptotic formula, is always $-\frac{5}{3}$. The same is true for the solutions $u_K$ of Equation~\eqref{eq:upper}; in particular, the coefficient of the lowest-order term, $\frac{\exp(z)-\sum_{j\leq K}z^j/j!}{z^K}$, has only a removable singularity at $0$ and no residue at $2$. It follows that the coefficients $l_n$ of $L(z)$ must also admit $l_n\sim c_{\infty} n^{2/3}2^{-n}$ for some constant $c_{\infty}$.\\
Multiplying by $n!(n-1)!\sim 2\pi \frac{n^{2n}}{e^{2n}}$, we have for $L_n$,
\begin{equation}
L_n\sim c_{\infty} \cdot \frac{n^{2n+2/3}}{2^{n}e^{2n}}
\end{equation}
which agrees with formula~\eqref{eq:kotesovec} with respect to exponential and polynomial terms (factoring in $(n+1)n\sim n^2$, since counting starts at $0$ instead of $1$ in \cite{kotesovec:litrees}).\\
The constant $c_{\infty}$ is not known exactly. Since the coefficients grow proportionally to $\frac{n^{2n+8/3}}{2^{n}e^{2n}}$ (when counting from $0$), one can estimate $c_{\infty}$ from the exact coefficients, yielding $c_{\infty}\approx 4.001655$ \cite{kotesovec:litrees}. 

%% file: restricted-degree-ode.tex
Incidentally, the solutions $s_K$ of Equation~\eqref{eq:lower} correspond to the (doubly-exponential) generating functions of $K$-ary labeled increasing trees, which we will denote by $L^{(K)}(z):=\sum_{n\geq 1}L^{(K)}_n (n!(n-1)!)^{-1}z^n$. The exponential and polynomial terms in the asymptotic formula for $L^{(K)}$ must remain the same (with one exception), and we are interested in how the constants $c_{K}$ compare to $c_{\infty}$. 
\subsection{$K=2$}
The simplest case is obtained by allowing an outdegree of at most $2$. This amounts to setting $K=2$, such that the summation in Equation~\eqref{eq:degf} is null. Naturally, $L^{(2)}(z)$ is the generating function of binary labeled increasing trees. Equation~\eqref{eq:degf} reduces to a functional equation,
\begin{equation}
L^{(2)}(z)=z+z/2L^{(2)}(z)
\end{equation}
with solution $L^{(2)}(z)=z/(1-z/2)$. Thus $K=2$ is the only case where the asymptotic formula $c\frac{n^{2/3}}{2^{n}}$ does not apply. The singularity at $2$ is a simple pole of degree $1$ and the coefficients are given by $l^{(2)}_n=2^{-n+1},n\geq 1$. Multiplying by $n!(n-1)!$, we unsurprisingly find that the number of binary labeled increasing trees equals $\frac{n!(n-1)!}{2^{n-1}}$ \cite{murthag:dendrograms,aldous:cladograms}.
\subsection{$K=3$}
If we allow up to $3$ leaves to merge at a time, we have $L^{(3)}(z)=s_3(z)$, so $L^{(3)}(z)$ solves Equations~\eqref{eq:2nddeg} and~\eqref{eq:2nddeg}. The latter, in fact, can be solved explicitly with the appropriate initial conditions, yielding
\begin{equation}
L^{(3)}(z)=\frac{2^{5/3}\exp(-(1/3)x)x}{(2-x)^{5/3}}
\end{equation}
The coefficient asymptotics are determined by the formula (see \cite{flajolet:combinatorics}, Ch.VI)
\begin{equation}
l^{(3)}_n\sim 2\exp\left(-\frac{2}{3}\right)\Gamma\left(\frac{5}{3}\right)^{-1}n^{\frac{2}{3}}
\end{equation}
We can again multiply the function coefficients $l^{(3)}_n$ by $n!(n-1)!$ to obtain the number of ternary labeled increasing trees. In order to make the asymptotic formula comparable to that of the number of all trees in \href{https://oeis.org/}{\texttt{OEIS}}, we again start counting from zero, meaning that we need to multiply by $n!(n+1)!$ instead. The number of ternary trees of size $n$ then follows the asymptotic formula 
\begin{equation}
L^{(3)}_{n}\sim 2\pi\exp\left(-\frac{2}{3}\right)\Gamma\left(\frac{5}{3}\right)^{-1}\frac{n^{2n+8/3}}{2^{n}e^{2n}}
\end{equation}
where the constant $c_3=2\pi\exp\left(-\frac{2}{3}\right)\Gamma\left(\frac{5}{3}\right)^{-1}$ is approximately $3.573428$.
\subsection{Almost all trees are quaternary, with the majority being ternary}
The asymptotic formulae for the number of ternary labeled increasing trees and the number of trees of arbitrary degree differ only up to their constants, with $c_3/c_{\infty}\approx 0.892982$. This shows that a randomly chosen labeled increasing tree has a probability of about $89\%$ of having only nodes of outdegree $2$ or $3$.\\
Going one step further, we investigate the class of quaternary labeled increasing tress. $L^{(4)}(z)=s_4(z)$ solves \eqref{eq:degf} truncated at $K=4$. The solution to the resulting differential equation does not admit a simple expression, but we know the exponential and polynomial factors present in the asymptotic expansion. By exact enumeration, we can determine the value of the constant factor $c_4$ to be approximately $3.977731$. Thus, for large $n$, trees with maximum node outdegree of $4$ already make up about $99.4\%$ of the set of all labeled increasing trees.\\

\begin{table}
\begin{center}
\begin{tabular}{ |r|r|r|r|r|}  
 \hline
 $n$& $L^{(2)}_n$& $L^{(3)}_n$& $L^{(4)}_n$ & $L_n$\\
 \hline
 1 & 1 & 1 & 1 & 1 \\
 2 & 1 & 1 & 1 & 1 \\
 3 & 3 & 4 & 4 & 4 \\
 4 & 18 & 28 & 29 & 29 \\
 5 & 180 & 320  & 335 & 336 \\
 6 & 2700 & 5360 & 5665 & 5687 \\
 7 & 56700 & 123760 & 131705 & 132294 \\
 8 & 1587600 & 3765440 & 4028430 & 4047969 \\
 9 & 57153600 & 145951680 & 156800490 & 157601068\\
 10 & 2571912000 & 7019678400 & 7567091700 & 7607093435\\                
 \hline
\end{tabular}
\end{center}
\caption{Counting sequences of binary, ternary, quarternary and general labeled increasing trees for sizes $1,\dots,10$. $L^{(2)}_n$, $L^{(3)}_n$ and $L^{(\infty)}_n$ have identifiers \texttt{A006472}, \texttt{A358072} and \texttt{A256006} on \href{https://oeis.org/}{\texttt{OEIS}}.}
\label{tab:cseq}
\end{table}

%% file: discussion.tex
The main aim of this work was to shed some light on the composition of Equation~\ref{eq:kotesovec}, describing the asymptotic growth of the number of labeled increasing trees. The results from from \cite{kotesovec:litrees} could be confirmed by our analysis. Indeed, using the existing techniques of analytic combinatorics to the treatment of differential equations \cite{flajolet:combinatorics}, it can be shown fairly straightforwardly how the exponential and polynomial factors in this formula arise. The constant $c_{\infty}\approx 4.001655$ is still not known exactly, however, with the remainder of the formula verified, one has the justification to determine it experimentally. On the other hand, the coefficient $c_{3}=2\pi\exp\left(-\frac{2}{3}\right)\Gamma\left(\frac{5}{3}\right)^{-1}$ appearing in the respective formula for ternary labeled increasing trees could be determined exactly.\\
We have noted before that the number of ternary and quaternary trees are quite large in comparison to the number of trees in total. The analysis reveals that asymptotically less than one percent of all trees has at least one node of degree larger than $4$. It appears that allowing nodes of higher outdegree reduces the number of nodes so quickly that in turn the contribution of trees with high arity to the total number of trees is relatively small. Notably, a similar observation is made in \cite{bondini:trees} regarding increasing Schr\"{o}der trees, even though the problem is viewed from a slightly different angle.\\
Due to the presence of the factor $n^{2n}$ in its asymptotic formula, the order of magnitude of the number $L_n$ of leaf-labeled increasing trees is bigger than that of permutations and even Bell numbers. It is also bigger by a factor of $\mathcal{O}(n!)$ than the number of increasing Schr\"{o}der trees. Interestingly, this is similar to the difference in coefficient growth between their binary restrictions, $L^{(2)}_n$ and $T^{(2)}_n$.\\
The combinatorial class $\mathcal{L}_n$ can play an important benchmarking role for algorithms making use of the theoretical foundations of the $\Lambda$-coalescents \cite{freund:coalstats,guindon:slfv,wirtz:slfvcoalrates}. Possible areas of research following up on the present analysis include the investigation of average tree height and balance, much like in \cite{bondini:trees}. Of particular interest appears to be the question which probability distributions can be induced on $\mathcal{L}_n$ (as well as $\mathcal{T}_n$) by the $\Lambda$-coalescents and other population-genetical or phylogenetical models, and will have to be addressed in future works.

%% file: main.bbl
\begin{thebibliography}{10}
\expandafter\ifx\csname url\endcsname\relax
  \def\url#1{\texttt{#1}}\fi
\expandafter\ifx\csname urlprefix\endcsname\relax\def\urlprefix{URL }\fi
\expandafter\ifx\csname href\endcsname\relax
  \def\href#1#2{#2} \def\path#1{#1}\fi

\bibitem{semple:phylo}
C.~Semple, M.~Steel, Phylogenetics, Oxford University Press, 2003.

\bibitem{gillespie:popgen}
J.~Gillespie, Population Genetics: A concise guide., John Hopkins, Baltimode,
  Md, 1998.

\bibitem{cannings:model}
C.~Cannings, The latent roots of certain markov chains arising in genetics: A
  new approach, ii. further haploid models, Advances in Applied Probability
  7~(2) (1975) 264--282.
\newblock \href {https://doi.org/10.2307/1426077} {\path{doi:10.2307/1426077}}.

\bibitem{moehle:coalescent}
M.~M\"{o}hle,
  \href{https://www.sciencedirect.com/science/article/pii/S0022519300920320}{Ancestral
  processes in population genetics—the coalescent}, Journal of Theoretical
  Biology 204~(4) (2000) 629--638.
\newblock \href {https://doi.org/https://doi.org/10.1006/jtbi.2000.2032}
  {\path{doi:10.1006/jtbi.2000.2032}}.

\bibitem{sagitov:coalescent}
S.~Sagitov, The general coalescent with asynchronous mergers of ancestral
  lines, J. Appl. Probab. 36 (1999) 1116--1125.

\bibitem{pitman:coalescent}
J.~Pitman, \href{https://doi.org/10.1214/aop/1022874819}{{Coalescents With
  Multiple Collisions}}, The Annals of Probability 27~(4) (1999) 1870--1902.
\newblock \href {https://doi.org/10.1214/aop/1022874819}
  {\path{doi:10.1214/aop/1022874819}}.

\bibitem{berestycki:beta}
J.~Berestycki, N.~Berestycki, J.~Schweinsberg, Beta-coalescents and continuous
  stable random trees, Ann. Probab. 35~(5) (2007) 18--35--1887.
\newblock \href {https://doi.org/10.1214/009117906000001114}
  {\path{doi:10.1214/009117906000001114}}.

\bibitem{barton:slfv}
N.~Barton, J.~Kelleher, A.~Etheridge, A new model for extinction and
  recolonization in two dimensions: quantifying phylogeography, Evolution
  64~(9) (2010) 2701--15.
\newblock \href {https://doi.org/10.1111/j.1558-5646.2010.01019.x.}
  {\path{doi:10.1111/j.1558-5646.2010.01019.x.}}

\bibitem{wakeley:coaltheory}
J.~Wakeley, Coalescent Theory, an Introduction, W. H. Freeman, 2008.

\bibitem{steel:shape}
M.~Steel, A.~McKenzie, The "shape" of phylogenies under simple random
  speciation models, Biol. Evol. Stat. Phys. 585 (05 2002).
\newblock \href {https://doi.org/10.1007/3-540-45692-9_9}
  {\path{doi:10.1007/3-540-45692-9_9}}.

\bibitem{wright:evolution}
S.~Wright, \href{http://symposium.cshlp.org/content/20/16.short}{Classification
  of the factors of evolution}, Cold Spring Harbor Symposia on Quantitative
  Biology 20 (1955) 16--24.
\newblock \href
  {\path{doi:10.1101/SQB.1955.020.01.004}}.

\bibitem{kingman:coalescent}
J.~F.~C. Kingman, On the genealogy of large populations, Journal of Applied
  Probability 19(A) (1982) 27--43.
\newblock \href {https://doi.org/10.2307/3213548} {\path{doi:10.2307/3213548}}.

\bibitem{aldous:balance}
D.~J. Aldous, Stochastic models and descriptive statistics for phylogenetic
  trees, from Yule to today, Statistical Science 16~(1) (2001) 23--34.

\bibitem{fischer:balance}
M.~Fischer, L.~Herbst, S.~Kersting, L.~K\"{u}hn, K.~Wicke,
  \href{https://arxiv.org/abs/2109.12281}{Tree balance indices: a comprehensive
  survey} (2021).
\newblock \href {https://doi.org/10.48550/ARXIV.2109.12281}
  {\path{doi:10.48550/ARXIV.2109.12281}}.

\bibitem{murthag:dendrograms}
F.~Murtagh, Counting dendrograms: A survey, Discrete Applied Mathematics 7~(2)
  (1984) 191 -- 199.
\newblock \href {https://doi.org/10.1016/0166-218X(84)90066-0}
  {\path{doi:10.1016/0166-218X(84)90066-0}}.

\bibitem{yule:process}
G.~U. Yule, A mathematical theory of evolution, based on the conclusions of Dr.
  J. C. Willis, f. r. s., Philosophical Transactions of the Royal Society of
  London B: Biological Sciences 213~(402-410) (1925) 21--87.
\newblock \href {https://doi.org/10.1098/rstb.1925.0002}
  {\path{doi:10.1098/rstb.1925.0002}}.

\bibitem{wirtz:emg}
J.~Wirtz, T.~Wiehe, The evolving moran genealogy, Theoretical Population
  Biology 130 (2019) 94--105.
\newblock \href {https://doi.org/10.1016/j.tpb.2019.07.005}
  {\path{doi:10.1016/j.tpb.2019.07.005}}.

\bibitem{flajolet:combinatorics}
P.~Flajolet, R.~Sedgewick, Analytic Combinatorics, 1st Edition, Cambridge
  University Press, New York, NY, USA, 2009.

\bibitem{aldous:probability}
D.~Aldous, R.~Pemantle, Random Discrete Structures, Springer IMA Volumes Math.
  Appl. 76, Berlin Heidelberg, 1996.

\bibitem{disanto:yulenodeimbalance}
F.~Disanto, A.~Schlizio, T.~Wiehe, Yule-generated trees constrained by node
  imbalance., Mathematical biosciences 246 1 (2013) 139--47.

\bibitem{disanto:external}
F.~Disanto, T.~Wiehe, Measuring the external branches of a Kingman tree: A
  discrete approach., Theoretical Population Biology 134 (2020) 92--105.

\bibitem{bondini:trees}
O.~Bodini, A.~Genitrini, C.~Mailler, M.~Naima., Strict monotonic trees arising
  from evolutionary processes: combinatorial and probabilistic study., Advances
  in Applied Mathematics 133 (2022) 102284.

\bibitem{aldous:cladograms}
D.~J. Aldous, Mixing time for a markov chain on cladograms, Combinatiorics,
  Probability and Computing 9~(3) (2000) 191--204.

\bibitem{wirtz:tld}
J.~Wirtz, M.~Rauscher, T.~Wiehe, Topological linkage disequilibrium calculated
  from coalescent genealogies, Theoretical Population Biology (2018).
\newblock \href {https://doi.org/10.1016/j.tpb.2018.09.001}
  {\path{doi:10.1016/j.tpb.2018.09.001}}.

\bibitem{frank:dendrograms}
O.~Frank, K.~Svensson, \href{https://doi.org/10.1080/00949658108810439}{On
  probability distributions of single-linkage dendrograms}, Journal of
  Statistical Computation and Simulation 12~(2) (1981) 121--131.
\newblock \href
  {\path{doi:10.1080/00949658108810439}}.

\bibitem{schroeder:trees}
E.~Schr\"{o}der, Vier combinatorische Probleme., Z. Math. Phys. 15 (1870)
  361--376.

\bibitem{oeis}
The on-line encyclopedia of integer sequences, \url{http://oeis.org} (2022).

\bibitem{kotesovec:litrees}
V.~Kotesovec, Sequence a256006, \url{https://oeis.org/A256006} (2015).

\bibitem{orrick:mergers}
W.~P. Orrick, How many ways to merge n companies into one big company - Bell or
  Catalan?, \url{https://math.stackexchange.com},  \url{https://oeis.org/A256006} (2016).

\bibitem{freund:coalstats}
F.~Freund, A.~Siri-J\'{e}gousse, The impact of genetic diversity statistics on
  model selection between coalescents, Computational Statistics and Data
  Analysis 156 (2020) 107055.
\newblock \href {https://doi.org/10.1016/j.csda.2020.107055}
  {\path{doi:10.1016/j.csda.2020.107055}}.

\bibitem{guindon:slfv}
S.~Guindon, H.~Guo, D.~Welch,
  \href{http://www.sciencedirect.com/science/article/pii/S0040580916300181}{Demographic
  inference under the coalescent in a spatial continuum}, Theoretical
  Population Biology 111 (2016) 43 -- 50.
\newblock \href {https://doi.org/https://doi.org/10.1016/j.tpb.2016.05.002}
  {\path{doi:10.1016/j.tpb.2016.05.002}}.

\bibitem{wirtz:slfvcoalrates}
J.~Wirtz, S.~Guindon,
  \href{https://www.sciencedirect.com/science/article/pii/S0040580922000314}{Rate
  of coalescence of lineage pairs in the spatial $\lambda$-fleming-viot
  process}, Theoretical Population Biology 146 (2022) 15--28.
\newblock \href {https://doi.org/https://doi.org/10.1016/j.tpb.2022.05.002}
  {\path{doi:10.1016/j.tpb.2022.05.002}}.

\end{thebibliography}
